\documentclass[twocolumn,aps,prd,nofootinbib,superscriptaddress]{revtex4-1}
\usepackage{graphics}
\usepackage{graphicx}
\usepackage{epsfig}
\usepackage[english]{babel}
\usepackage{amsmath} 
\usepackage{verbatim} 
\usepackage{mathrsfs}
\usepackage{amsfonts}
\usepackage{amssymb}
\usepackage[latin1]{inputenc}
\usepackage{hyperref}

\newcommand{\beq}{\begin{equation}}
\newcommand{\eeq}{\end{equation}}
\newcommand{\barr}{\begin{eqnarray}}
\newcommand{\earr}{\end{eqnarray}}
\newcommand{\bea}{\begin{eqnarray*}}
\newcommand{\eea}{\end{eqnarray*}}

\begin{document}

\title{What do we talk about when we speak of cosmological redshift?}

\author{Gabriel R. Bengochea}
\email{gabriel@iafe.uba.ar} \affiliation{Instituto de Astronom\'\i
a y F\'\i sica del Espacio (IAFE), CONICET - Universidad de Buenos Aires, (1428) Buenos Aires, Argentina}

\begin{abstract}

From the first observations made by Slipher, our understanding and interpretation of the cosmological redshift was evolving until reaching the current consensus, through the expanding universe and the emergence of modern physical cosmology within the framework of General Relativity. The redshift is one of the most basic concepts of astronomy, and is one of the few observational parameters that can be measured directly. To refer to the temporal evolution of objects or cosmic structures in the universe, we often do so indistinctly through cosmic time or cosmological redshift. But repeatedly this connection ends up generating confusion not only among popular science communicators but also within the professional astronomical community. In this article, we will make a pedagogical approach to the link between cosmic time and cosmological redshift, and we will also clarify several common misunderstandings around this relation.

\end{abstract}

\maketitle

\section{Introduction}
\label{intro}

The Big Bang is our best model about how we think the universe works, and the discovery that distant galaxies have recession velocities proportional to their distances is the cornerstone of modern cosmology. Cosmological redshifts are now well understood within the framework of Einstein's theory of General Relativity \cite{einstein15}. However, there are still misunderstandings about the concepts of expansion and redshift, not only among popular science communicators but also within the professional astronomical community \cite{ellis93, DL2003, tdavis03, sciam2005, Francis07, Lewis16}. Sometimes, when the expansion of the universe is involved, misinterpreted statements concerning redshift arise.

Being a fundamental observational parameter that can be obtained directly with a measuring instrument, the redshift (denoted by $z$) is one of the most basic concepts of astronomy. The observation and recording of the spectral lines in galaxies undoubtedly reveal this phenomenon.

The absorption lines in the spectrum of a galaxy can be used to obtain information about chemical elements present in that galaxy. Each chemical element generates a different pattern of absorption lines in the spectrum, at wavelengths that can be measured extremely reliably by spectrographs. When we identify some lines of specific chemical elements in the spectra of the galaxies, and compare them with the lines of spectra in experiments carried out in a laboratory, we unequivocally find that the patterns of the spectra of the galaxies are the same, but they are shifted with respect to those of the terrestrial laboratory. The most spectra are displaced towards the red color, and therefore we refer to this phenomenon as \emph{redshift}.

The first records of redshifts in galaxies were obtained by the astronomer Slipher \cite{slipher1, slipher2, slipher3}. Later, Wirtz and Lundmark (e.g. \cite{wirtz, lundmark}) mention the existence of spiral \emph{nebulae} whose redshifts seemed to increase with distance. However, there was still no clear relation between redshift and distance. Until Hubble discovered Cepheid variable stars in the Andromeda nebula \cite{hubble25}, it was only possible to infer relative distances. Cepheids allowed Hubble to estimate a true distance to Andromeda. Since then, we also know that those nebulae are actually galaxies more or less similar to ours, the Milky Way.

In the times of Lema\^itre and Hubble, redshifts (and blueshifts) were interpreted as a Doppler effect. In Doppler effects, redshifts are a consequence of velocities involved between sources and observers. First Lema\^itre \cite{lemaitre27}, and then Hubble \cite{hubble29}, obtained velocities of a few galaxies by using a linear velocity-distance law. In particular, Hubble took the radial velocities for 24 galaxies with 'known' distances and fitted them to certain relation (now known as the Hubble law), obtaining a high value (similar to that of Lema\^itre in 1927) for what we now call the Hubble constant. By the early 1930s, Hubble had measured redshifts $z\simeq0.02$, and then a linear relation between redshift and distance was becoming clearer. The conclusion was (a little later) that the universe is expanding.

Theorists almost immediately realized that these observations could be explained by redshifts that appear in certain cosmological solutions to Einstein equations of General Relativity.

In a recent work \cite{paturel17}, the authors describe some interpretations of the Hubble law and it is remembered that the first suggestion for a cosmological redshift was from W. de Sitter, as part of a static solution of Einstein equations \cite{desitter1, desitter2, desitter3}. In fact, in \cite{eddington} Eddington mentioned that, within the de Sitter model, the displacement of spectral lines observed could be explained by a slowing down of atomic vibrations, and that it would be wrongly interpreted as a motion of recession. Hubble himself in his renowned work \cite{hubble29} writes that a possible explanation for the distance-redshift law could be due to the de Sitter effect. Humason \cite{humason} is another author who mentions that Hubble's observational results could have something to do with the de Sitter effect.

Adopting Einstein's Cosmological Principle, that is, under the assumption of isotropy and spatial homogeneity on large scales, Friedmann, Lema\^itre, Robertson and Walker (FLRW) found solutions to Einstein field equations that contemplate expanding universes \cite{friedmann, lemaitre27, robertson1, robertson2, walker}. Our modern cosmology is based on these FLRW models, and these solutions could be used to give a more elaborated theoretical sustenance to the ideas and observations of the pioneer astronomers. In fact, today we think that the most correct interpretation of redshift is that which involves an expanding universe and not through a Doppler effect, as originally thought by astronomers such as Slipher, Lundmark and Hubble when they used the equation $V=cz$ to calculate velocities.\footnote{See for instance \cite{hubble31}, where Hubble and Humason make reference to that if actual velocities of recession are involved, a correction to the equation should be made, making an allusion to the relativistic Doppler effect.}

Within the General Relativity framework, the cosmological redshifts arise since the proper distances between comoving objects increase with time. But then, the velocities generated are dominantly due to the expansion of space, determined by the cosmological model chosen to describe the universe, and not due to \emph{peculiar} (local) velocities through space. In addition, as we consider increasingly distant objects, peculiar velocities of distant galaxies becomes negligible with respect to the velocity of expansion of the universe at the location of such a galaxy. Therefore, although some popular literature often uses the expression 'Doppler redshift' instead of cosmological redshift, it cannot be calculated with the Doppler equation, as already explained by various authors \cite{harrison81, harrison93, ellis93, kiang03, DL2003, tdavis03, Francis07}.

Redshift $z$ is generally defined as the change registered between the frequency that light had at the time of emission from an object, $\nu_{\rm{em}}$, and the frequency observed today in a detector, $\nu_{\rm{obs}}$. Now, we know that considering a light ray coming to us from a distant galaxy along the radial direction, and traveling through a null geodesic of the FLRW metric, by a simple cinematic analysis we find that the light in its journey must change its frequency. In this manner, the cosmological redshift $z$ turns out to be the quotient between the value of the scale factor of the universe today $a_0$ and that corresponding to the time $t$ of the emission $a(t)$ \cite{weinberg}. This is,
\beq\label{redshift}
1+z\equiv \frac{\nu_{\rm{em}}}{\nu_{\rm{obs}}}=\frac{a_0}{a(t)}
\eeq
Then, differentiating Eq. \eqref{redshift} with respect to $t$ is obtained
\beq\label{maineq}
dt=-\frac{dz}{H(z) (1+z)}
\eeq
where redshift is used instead of time to write the Hubble parameter $H(z)$. This last equation and its misinterpretations are the main focus of this article.

As it was mentioned above, redshift is one of the most fundamental observational data that can be obtained directly with a measuring instrument. Quantities such as velocities are not directly observable, therefore, we have never directly measured recession velocities. Apart from redshift, for almost everything else one must adopt a theory of gravity, a cosmology with certain cosmological parameters, and then do the calculations to obtain recession velocities, the age of the universe, and other magnitudes of interest .

Because of the univocal relationship between redshift $z$ and time $t$ in Eq. \eqref{maineq}, we often speak of events happening at a given redshift instead of at a given time. This is convenient because the redshift is observable and usually has a great effect on the rates of physical processes. For instance, we mean that the universe would have started to accelerate at, say, $z\simeq0.6$, or that the decoupling of matter and radiation took place at $z\simeq1100$, when the universe was very young. Misinterpretations appear in other circumstances, when one needs to be more specific about what $z$ and $t$ one is referring to in that relation. Surprisingly, many people think that an object that we observe today at a given $z$, at an earlier time of its evolution the same object was in a very (higher) different $z$.\footnote{Due to the deceleration or acceleration of the universe, the redshift $z$ of a galaxy is not fixed but changes a small amount $\Delta z\simeq 10^{-8}$ in about 100 years. This effect is known as \emph{cosmological redshift drift} \cite{sandage62, loeb98}, and in the future it may be used to directly measure the expansion rate of the universe. However, here we will not take it into account for two reasons: for being a small effect and because it is not relevant to the conceptual discussion that this article intends.}

We usually see numerical simulations of stellar objects or about large scale structures of the universe, where we are shown \emph{snapshots} labeled with different values of cosmological redshift. Clearly, an univocal association is suggested, through the relation \eqref{maineq}, with specific times for the formation and the temporal evolution of the structure shown in such simulations. But then, it is usual for many people to interpret that these simulations are showing the \emph{same object} passing through \emph{different redshifts} throughout its evolution, and that this is what happens in the real universe, according to our expanding universe paradigm in the framework of General Relativity.

When we read about a supermassive black hole being detected in a quasar at $z=7.54$, adopting the current concordance cosmology \cite{planck15}, it is mentioned then that the quasar is situated at a cosmic age of just $t_{\rm{age}}=\rm{690\:Myr}$ after the Big Bang \cite{Banados2017}. But what do we mean when we mentioned that at redshift $z$ the age of the universe was $t_{\rm{age}}$? Only the objects that we see today with that specific $z$ were situated at a cosmic age of $t_{\rm{age}}$? The answer is no. What then is the meaning of 'univocal relation' in the equation for $t-z$?

In this article, we will shed light on these typical misinterpretations with a pedagogical approach through spacetime diagrams. In Sect. \ref{sectdos} we will describe some basic concepts about the standard cosmological model, in Sect. \ref{sectres} we will present some typical misconceptions about the $t-z$ relation, and finally, in Sect. \ref{conclusiones} we will present some final comments. Sometimes, to make the description of a concept, I will use generically the term "galaxy" to describe any object located at a given redshift $z$.

\section{Brief review of general concepts}
\label{sectdos}

In this section, we will summarize some basic concepts and definitions regarding the concordance cosmological model in the framework of General Relativity, and we will make some general comments that will serve as a basis to understand the following section. The reader interested in more details, can refer for instance to \cite{weinberg, kolbturner, peacock, ellis93}.

The starting point of the standard cosmological model is to assume that the universe is spatially homogeneous and isotropic on large scales, and then the spacetime can be well described by the Friedmann-Lema\^itre-Robertson-Walker (FLRW) metric. This metric can be written as \cite{muk05},
\beq\label{flrwmetric}
ds^2=-c^2dt^2+a^2(t)\:[d\chi^2+f^2(\chi)d\Omega^2]
\eeq
where $dt$ and $d\chi$ are the time and comoving coordinate separations respectively, and where $d\Omega^2=d\theta^2+\sin^2(\theta)d\varphi^2$ is the angular part with $\theta$ and $\varphi$ being the angles in spherical coordinates. The scale factor $a(t)$ has dimensions of distance, and the function $f(\chi)=\sin\chi$, $\chi$ or $\sinh \chi$ for closed ($k=+1$), flat ($k=0$) or open ($k=-1$) geometries respectively. Observers with \emph{constant} comoving coordinate $\chi$ and who see an isotropic and homogeneous universe are known as \emph{comoving observers}. They are the ones who simply follow the Hubble flow\footnote{Note that, since we have non-zero local peculiar velocities, we are not comoving observers. Therefore, in general, times often shown in the literature (such as the age of the universe, for example) are not, strictly speaking, what a clock on Earth measures.}.

Defining the Hubble parameter as $H(t)\equiv \frac{\dot{a}(t)}{a(t)}$, and substituting Eq. \eqref{flrwmetric} into Einstein equations, the 0-0 component of those equations tell us that the evolution of the scale factor $a(t)$ is determined by the composition of the universe, according to the Friedmann equation:
\beq\label{friedeq}
H^2(t)=\frac{8\pi G}{3c^2}\rho(t)-\frac{kc^2}{a^2(t)}
\eeq
where $\rho(t)$ is the total density of the cosmological fluid (radiation, matter, dark energy, etc), and a \emph{dot} will denote derivatives with respect to the time $t$. Densities can be normalized to the present critical density $\rho_{\rm{crit}}=3c^2H_0^2/8\pi G$, and then if we consider a universe composed only of matter (baryons plus dark matter) and dark energy, we can write these contributions as $\Omega_{\rm{m}}=\rho_{\rm{m}}/\rho_{\rm{crit}}$ and $\Omega_{\rm{\Lambda}}=\rho_{\rm{\Lambda}}/\rho_{\rm{crit}}$ respectively. We could also assume here, without loss of generality, the observationally favored flat case ($k=0$), so that $\Omega_{\rm{m}}+\Omega_{\rm{\Lambda}}=1$. Thus, solving the conservation equation for the $\rho_i(z)$, Eq. \eqref{friedeq} can be written as
\beq\label{friedeq2}
H(z)=H_0\Big[\Omega_m (1+z)^3+\Omega_\Lambda \Big]^{1/2}
\eeq
where $H_0$ is the Hubble constant and, for numerical purposes, we will adopt typical values fixing $\Omega_{\rm{m}}=0.3$, $\Omega_{\rm{\Lambda}}=0.7$ and $H_0=70 \: {\rm km\:s^{-1}Mpc^{-1}}$.

The time $t$ is the proper time that a comoving observer measures, and sometimes it is called cosmic time. This is the time that appears in the FLRW metric and the Friedmann equation.

On the other hand, the proper (radial) distance, $D(t) = a(t)\chi$, is defined as the changing distance (with $dt = 0=d\Omega$) between us and an object with comoving coordinate $\chi$. Thus, this distance increases (or decreases) with the scale factor $a(t)$. The information about if $a(t)$ is increasing, decreasing or constant comes from astronomical observations, and these ones tell us that there is a (red)shift in the spectral lines from distant galaxies. Then, under the hypothesis of expansion of the universe, the proper distances $D(t)$ between distant galaxies (located in fixed comoving positions $\chi$) must be increasing with time, because the scale factor $a(t)$ is growing. If $a(t)$ is increasing, then there will be a redshift in frequency given by $a_0/a(t)$, conventionally denoted by $1+z$ and shown in Eq. \eqref{redshift}.

We will denote $\chi(z)$ as the fixed comoving coordinate of a object observed today at redshift $z$. Galaxies are not necessarily in fixed positions, but the $\chi(z)$ coordinates could be changing in time due to velocities produced by gravitational effects between neighboring objects. So, the total velocity of an object is defined as,
\beq\label{Ddot}
\dot{D}(t,z)\equiv V(t,z)=\dot{a}(t)\: \chi(t,z)+a(t)\:\dot{\chi}(t,z)
\eeq
In the last equation, the second term, $a\:\dot{\chi}$, is the peculiar velocity of the object, $v_{\rm{pec}}$. Since cosmology deals with large-scale structure (large distances), and we know that non-relativistic matter (and photons too) has momentum decreasing as $p\propto 1/a$ \cite{weinberg}, peculiar velocities are considered negligible with respect to the recession velocity shown in the first term of \eqref{Ddot}. Therefore, in cosmology is usual to fix $\dot{\chi}=0$ (objects will be in fix $\chi(z)$ positions), and then we write the first term for the recession velocity as $v_{\rm{rec}}(t,z)=\dot{a}(t)\: \chi(z)$. With the definition of the Hubble parameter, the equation for $v_{\rm{rec}}$ gives the famous Hubble law, $v_{\rm{rec}}(t) = H(t)D(t)$. Notice that the theoretically predicted linear velocity-distance relation $V=HD$, can exist only if the matter distribution is uniform. Remember that peculiar velocities of massive objects correspond to local velocities (hence $v_{\rm{pec}}< c$), and are responsible for Doppler effects. In contrast, recession velocities can be arbitrarily large because they are due to the expansion of space.

Let us emphasize something more about velocities. While in the case of Special Relativity local velocities such as $v_{\rm{pec}}$ depend only on $z$, in the General Relativity description, velocities as the shown in Eq. \eqref{Ddot}, have an additional dependence with time $t$. The value of velocity $v_{\rm{rec}}$, in the same spatial position and with the same redshift, is changing due to the expansion of the universe. Then, in cosmology we also have to choose when we want to calculate velocities. Since the first order approximation $V\approx cz$ (typically used in the Hubble law) is shared by both, Special and General Relativity, it used to be used to calculate the velocities of galaxies. But this simple equation is only valid for $z\lesssim 0.3$ (See for instance the discussion in \cite{DL2003, tdavis03}).

The times in which the objects emitted the light that we see today are those that participate in the definition of our past light cone. Since light rays travel along null geodesics, setting $ds = 0$ in the FLRW metric, radial comoving distances will result simply from solving $c\:dt=a(t) d\chi$. Then, by integrating this last equation it results the coordinate of an object on the past light cone. But as we have seen, Eq. \eqref{maineq} allows us to use the redshift instead of time and thus the comoving coordinate of an object can be written as,
\beq\label{cono2}
\chi(z)=\frac{c}{a_0}\:\int_0^{z} \frac{dz'}{H(z')}
\eeq
where $z=0$ corresponds to today at $t_0$. One can see that, contrary to the case of a Doppler effect, the redshift of a given galaxy has not so much to do with its velocity, but rather with its position. This is an important concept for what we will discuss in the next section. Finally, when it is necessary to know recession velocities, we simply have to use Eq. \eqref{cono2} into $v_{\rm{rec}}(t,z)=\dot{a}(t)\: \chi(z)$. As it was mentioned, velocities are not observed directly but must be calculated with an assumed cosmological model.

\section{Redshift confusions}
\label{sectres}

In this section, we will see all those concepts addressed in the previous sections through spacetime diagrams. After that, we will analyze some misunderstandings.
\begin{figure*}
\begin{center}
\includegraphics[scale=0.4]{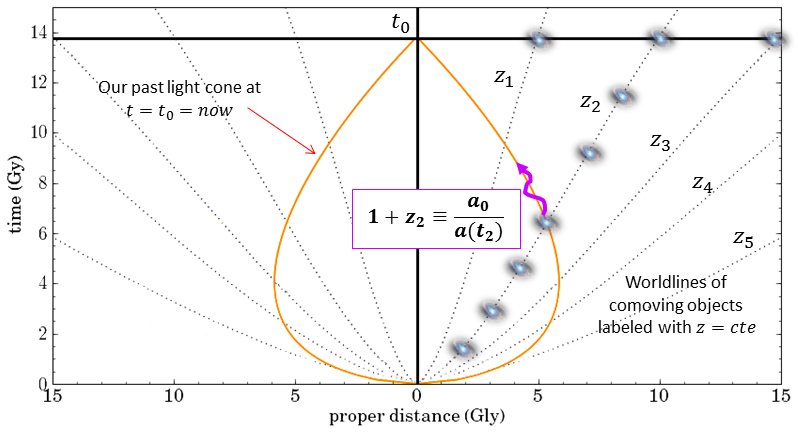}
\includegraphics[scale=0.4]{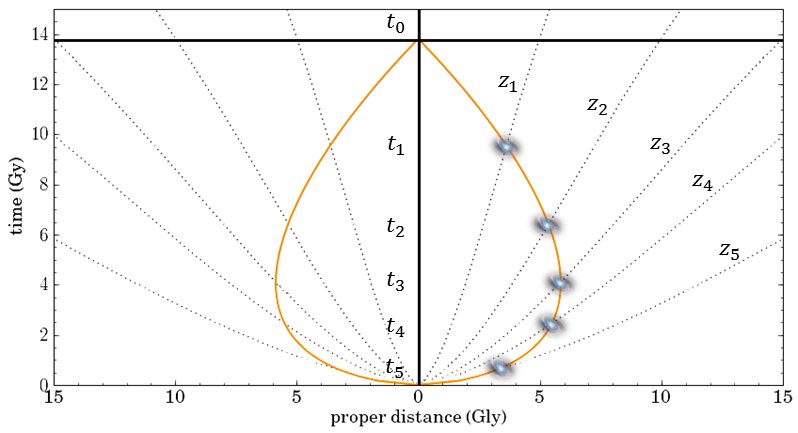}
\end{center}
\caption{Spacetime diagrams $t$ vs proper distance $D$ based on the FLRW metric, adopting the concordance $\Lambda$CDM model and where we are using to plot the diagrams that $D(t)=a(t)\chi$. Left:  Worldlines of comoving objects are shown with dotted lines, and labeled with different redshifts $z_i$. Our past light cone is shown with a solid line, resulting from $ds=0=d\Omega$ in the FLRW metric. In this figure we can also see how we define the redshift of a galaxy at $z=z_2$ by light rays emitted on our past light cone, at time $t_2$, when the scale factor was $a(t_2)$. Right: Galaxies on our past light cone indicate the points associated with the relation $t-z$ of Eq. \eqref{maineq}. Different values of $z$ imply different galaxies.}
\label{figred}
\end{figure*}

Let us start with basic concepts such as redshift, emission time and comoving objects. Figure \ref{figred} shows a spacetime plot (i.e. cosmic time $t$ versus proper distance $D$), where a FLRW metric is assumed. Our comoving coordinate is the central vertical worldline, and dotted lines show the worldlines of comoving objects. Notice that the changing recession velocity of a comoving object is reflected in the changing slope of its worldline. Redshifts of the comoving galaxies appear labeled on each comoving worldline as $z_i$. Our current past light cone is shown with a solid line and it delimits the events in the universe that we can currently see. These spacetime diagrams also assume the observationally favored flat $\Lambda$CDM concordance model, where we have used a Hubble constant $H_0=70\: {\rm km\:s^{-1}Mpc^{-1}}$ and normalized matter and dark energy densities given by $\Omega_{\rm{m}}=0.3$ and $\Omega_{\rm{\Lambda}}=0.7$ respectively.

Figure \ref{figred} (Left), shows the particular "trajectory" of a galaxy at $z_2$. That redshift is defined from the ray of light that we receive today from this galaxy, when the scale factor is $a_0$, but that was emitted at time $t_2$ when the same galaxy crossed our current past light cone, and the scale factor was $a(t_2)$. The expansion of the universe means that the proper distance $D$ of this galaxy increases, because the scale factor grows (determined by the temporal changes of $H_0$ and the densities of the content of the universe, i.e. $\Omega_{\rm{m}}$ and $\Omega_{\rm{\Lambda}}$). But note that the $z$ value of the galaxy does not change in time and is fixed throughout its evolution, since the redshift is associated with a fixed comoving position (see footnote 2). It is also shown the theoretical prediction for the location and distance of that galaxy at the current time $t_0$ (and other two to $z_1$ and $z_3$), according to the paradigm of expansion of an isotropic and homogeneous universe with matter and dark energy.

In Fig. \ref{figred} (Right), it is shown some galaxies crossing our current past light cone, indicating the points that are associated with the relation $t-z$ of the Eq. \eqref{maineq}. Notice that involving different values of $z$ means to deal with \emph{different} galaxies. Times $t_i$ associated with the $z_i$ in the Eq. \eqref{maineq} are then the \emph{emission times}, and correspond to the instants in which the galaxies "cross" by our past light cone.

\begin{figure}
\begin{center}
\includegraphics[scale=0.4]{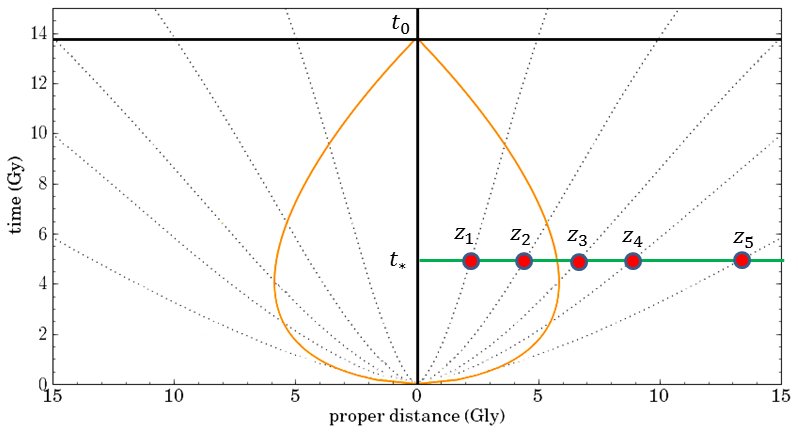}
\end{center}
\caption{At any time $t_*$, when the age of the universe is $t_{\rm{age}}$, there are many galaxies (filled dots) with the same time of evolution and with all possible values of different redshifts.}
\label{interp1}
\end{figure}

Let us analyze now some of the misunderstandings that usually appear referring to these concepts. Thinking of a certain redshift $z$ as an indicator of a time of evolution can lead to misinterpretations. Figure \ref{interp1} shows that in reality at any time $t_*$, when the age of the universe is $t_{\rm{age}}$, there are many galaxies with the same time of evolution and with all possible values of different redshifts. For instance, when we say that 200 million years after the Big Bang the first galaxies were formed, they did it to all redshifts $z$ at the same time.
\begin{figure*}
\begin{center}
\includegraphics[scale=0.21]{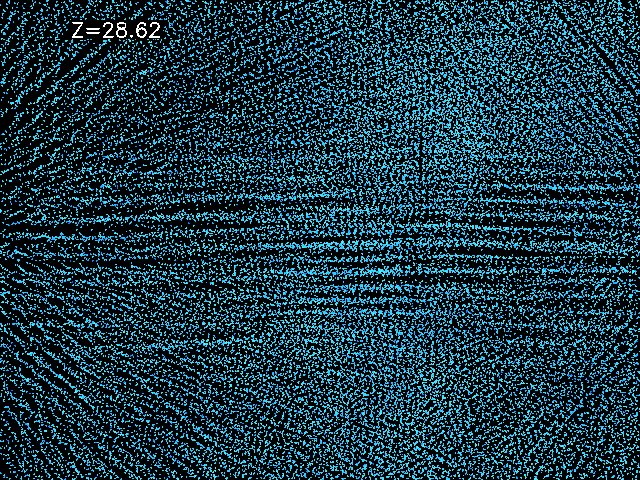}
\includegraphics[scale=0.21]{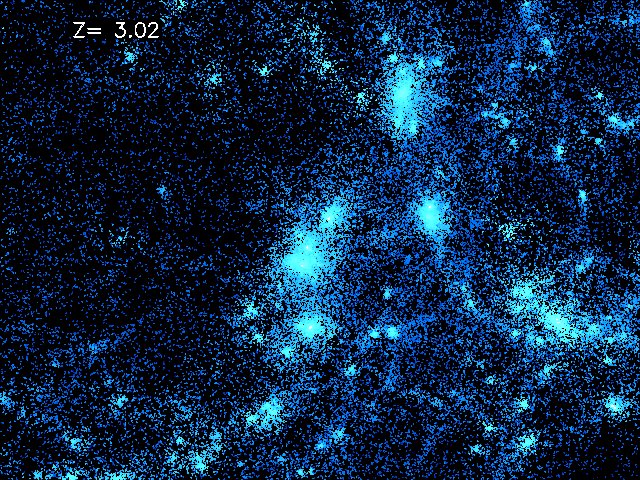}
\includegraphics[scale=0.21]{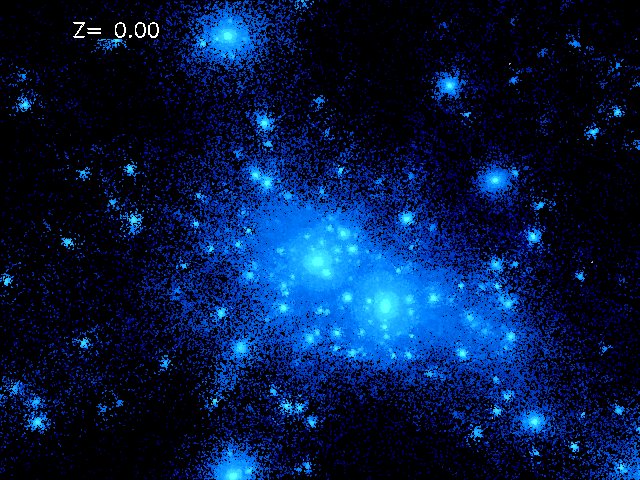}
\end{center}
\caption{Formation of a group of galaxies quite similar to our Local Group. The region shown here is about 4 Mpc in size. The center of the field of view is fixed in the same comoving position, tracking the progenitor of the group.  Frames shown at three different redshifts: $z=28.62$, $z=3.02$ and $z=0$. Simulations were performed at the National Center for Supercomputer Applications, by Andrey Kravtsov (The University of Chicago) and Anatoly Klypin (New Mexico State University), \url{http://cosmicweb.uchicago.edu/filaments.html}}
\label{grupogal}
\end{figure*}
\begin{figure*}
\begin{center}
\includegraphics[scale=0.18]{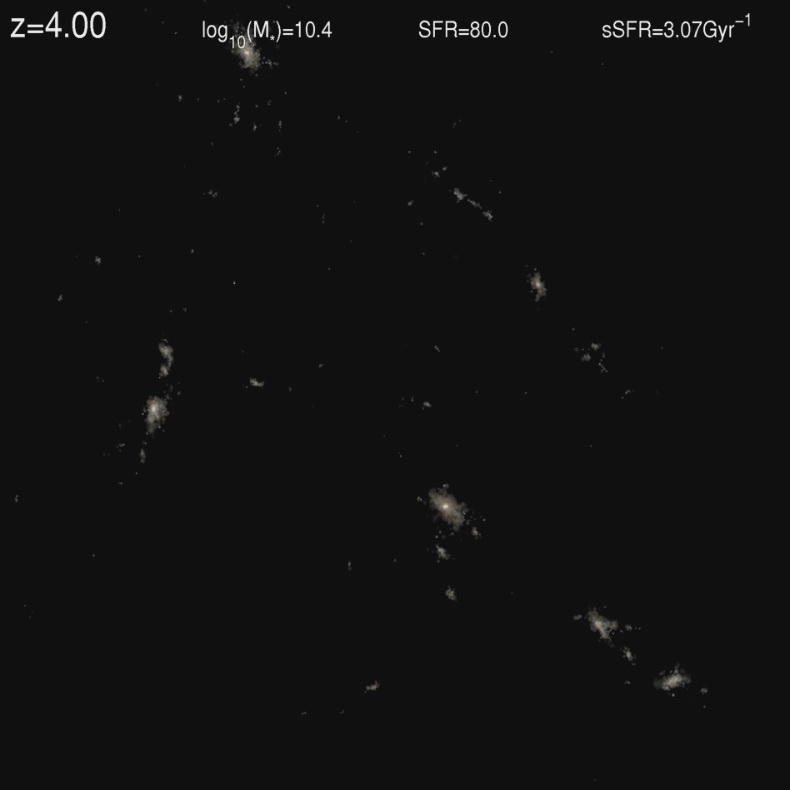}
\includegraphics[scale=0.18]{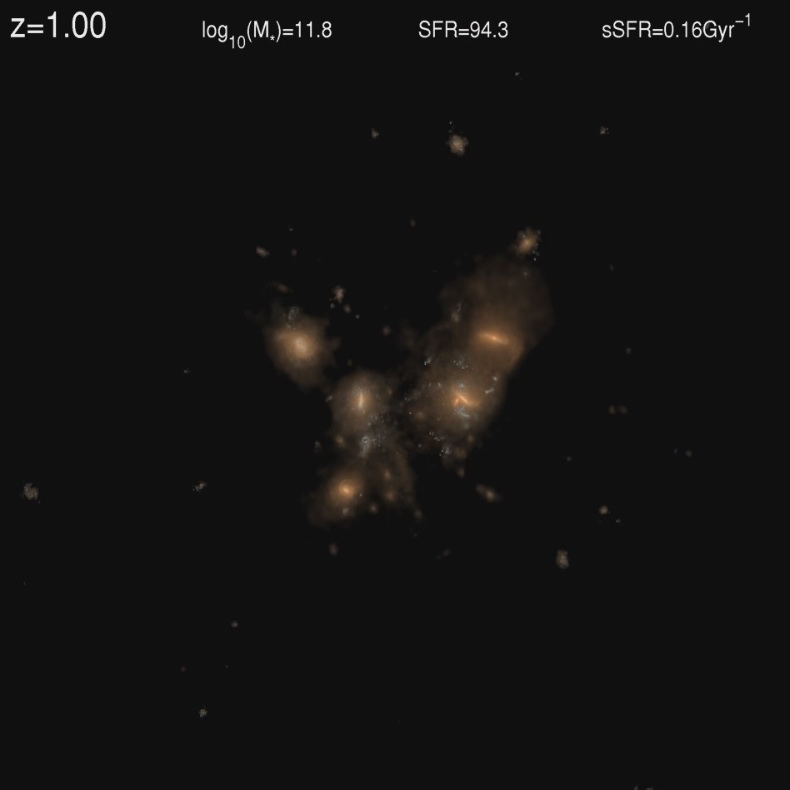}
\includegraphics[scale=0.18]{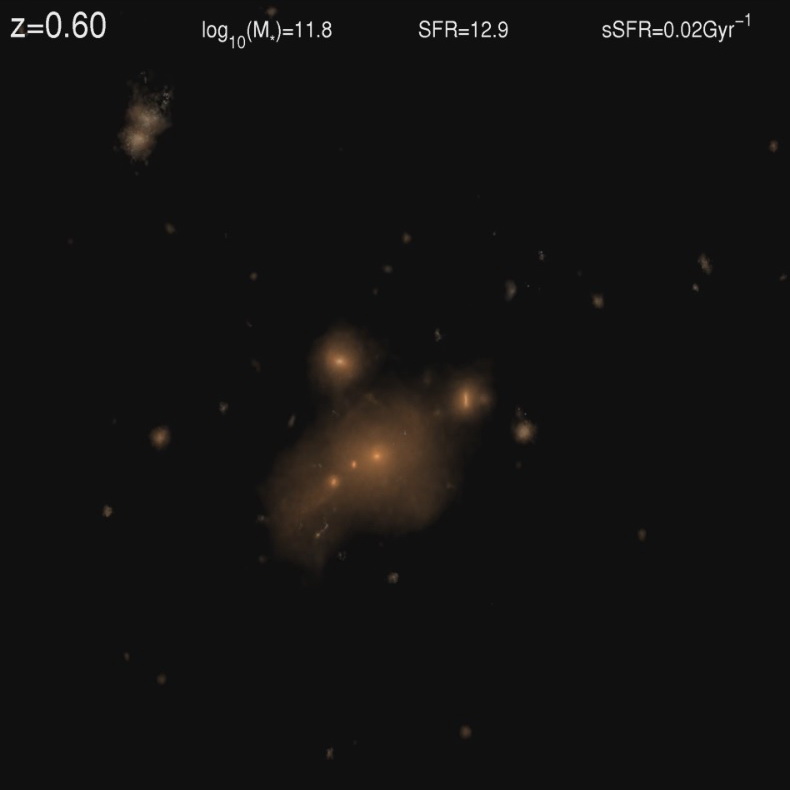}
\includegraphics[scale=0.18]{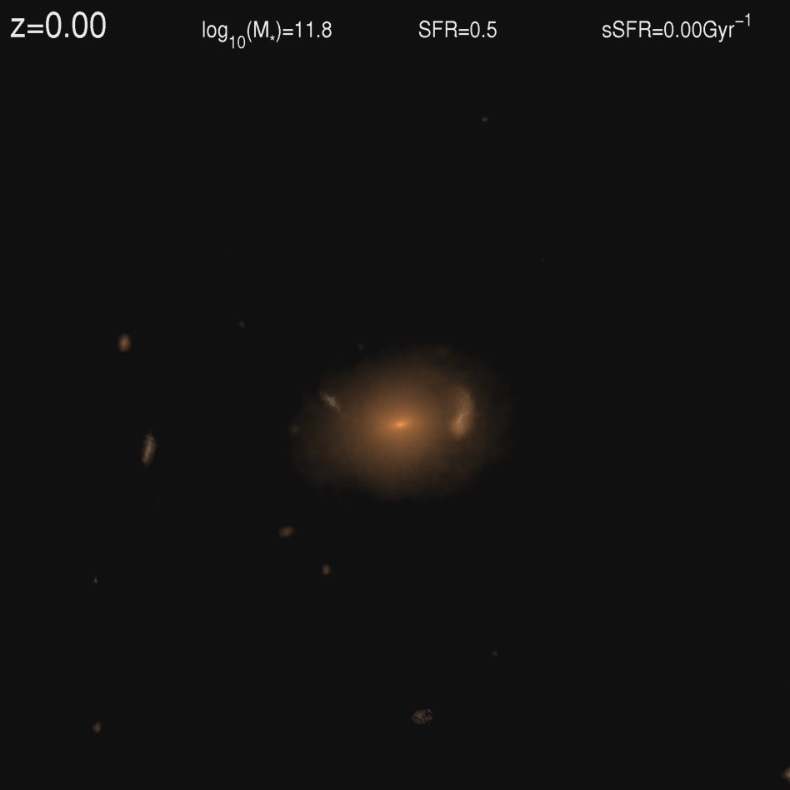}
\end{center}
\caption{Time evolution from redshift $z=4$ to $z=0$, demonstrating the formation of a massive elliptical galaxy as a result of a multiple merger around $z\sim1$. Snapshots show stellar light in a region of 1 Mpc on a side. When viewing these simulations, one should bear in mind that in the real universe the same object (or comoving position) does not take different values of cosmological redshift during its evolution. Simulations performed by \emph{Illustris Collaboration}, \url{http://www.illustris-project.org/media}}
\label{illustris}
\end{figure*}

In numerical simulations of formation and evolution of objects or structures, it is typical to show these evolutions by \emph{snapshots} labeled with different cosmological redshifts since that, as we have seen, Eq. \eqref{maineq} allows us to relate $t$ with $z$. For instance, in Fig. \ref{grupogal} are shown three sequences of the formation of a group of galaxies, similar to our Local Group. In general, in these simulations the center of the field of view is fixed in the same comoving position, tracking the progenitor, and showing the evolution of the same region. A non-specialist reader might then ask: how does the same region, in the same comoving position, have different values of cosmological redshifts during its evolution?

As a result of this confusion, many people misinterpret what the simulations are showing. Figure \ref{illustris} shows three snapshots of the time evolution of a comoving region of 1 Mpc on a side and from redshift $z=4$ to $z=0$, demonstrating the formation of a massive elliptical galaxy as a result of a multiple merger around $z\sim1$. Another example can be seen in Fig. 2 of \cite{dimatteo12}, where is shown through an excellent cosmological simulation the temporal evolution of first quasars in the universe from $z=8$ to $z=4.75$. But it is clear that at the time when the quasars start to form (say at the time associated with the redshift $z = 8$ shown in that simulation), actually at that time all quasars start to form at all the values of redshift and not only at $z = 8$.

\begin{figure*}
\begin{center}
\includegraphics[scale=0.4]{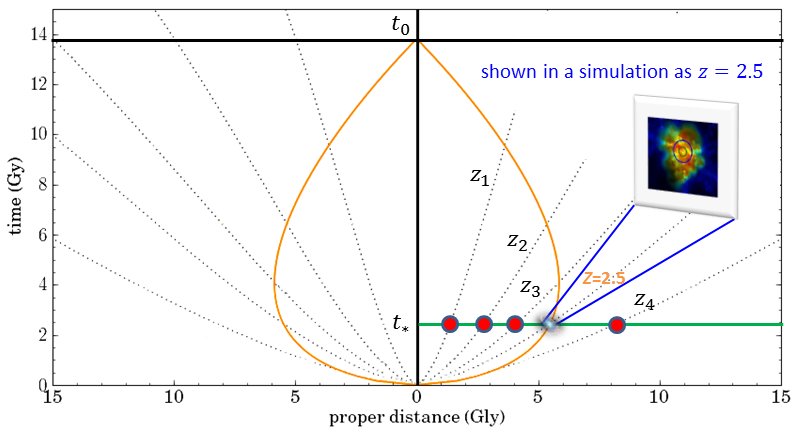}
\includegraphics[scale=0.4]{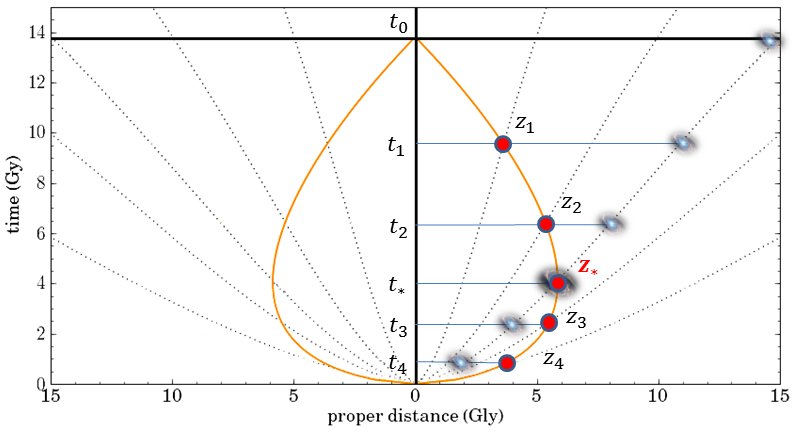}
\end{center}
\caption{Left: In the real universe, there are many objects (filled red dots) located at different $z$ with the same evolution time $t_*$. The only peculiarity of a redshift value shown in a simulation (for example at $z = 2.5$), is that at the moment of sending us the light that reaches us today, the object was on our past light cone. Right: Numerical simulations labeled with $z$ are showing us dots $t-z$ on our past light cone. To show the time evolution of the same galaxy (as the one at $z_*$) labeling it with different values of $z$, we are also assuming that all the galaxies evolved in an identical manner, and in that way we can associate the galaxy followed with each point on the past light cone.}
\label{interp2}
\end{figure*}

One must keep in mind that in the real universe, the same object (or comoving position) does not take different values of cosmological redshift during its evolution, and that the equation $t-z$ relates only the points on our past light cone, for different galaxies. In Fig. \ref{interp2} (Left) we can observe that the only peculiarity of the redshifts shown in a simulation, for example at $z = 2.5$, is that the object at the moment of sending us the light that reaches us today, it was on our past light cone at the time of evolution $t_*$. But, again, note that with the same evolution time there are many other objects located at different $z$, indicated with filled red dots on the same figure.

To the right of the Fig. \ref{interp2}, it is shown in the diagram how a simulation about the temporal evolution of the \emph{same} galaxy (located at a fixed $z_*$) should be interpreted correctly. The evolution and expansion of the universe take away the galaxy from us, growing its proper distance $D$ to us, but staying in a fixed comoving position $\chi$. To be able to label its evolution with different values of $z$ we must also suppose that at a given time all the galaxies evolved in an identical manner, and therefore we can associate to that galaxy the $z$ of \emph{other} galaxies (in the figure: $z_1$, $z_2$, $z_*$, $z_3$ and $z_4$) at the time of crossing our past light cone. Note that only at time $t_*$ the redshift $z_*$ is the one that really corresponds to the galaxy whose evolution we are following in the example.

\begin{figure}
\begin{center}
\includegraphics[scale=0.4]{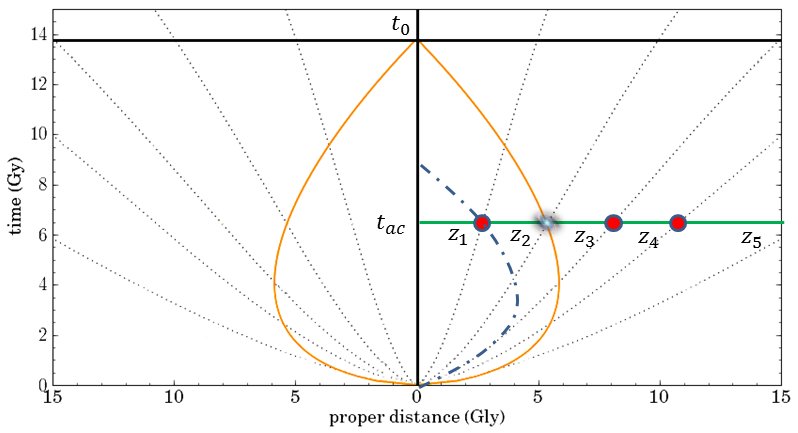}
\end{center}
\caption{For a same cosmic time $t$ there are many values of $z$. In this example, we can see that when the universe starts to accelerate at time $t_{\rm{ac}}$, that event today has associated the redshift $z_2$. But that same time $t_{\rm{ac}}$ has also associated other redshift values, some of them shown with filled red dots. An observer in the past, whose past light cone is drawn with a dash-dotted line, calculates that the acceleration of the universe occurred at the same $t_{\rm{ac}}$ but at redshift $z_1$.}
\label{interp4}
\end{figure}

As a last case, we will now consider in these spacetime diagrams what a given physical event means, at a given time, for different values of redshift. Let us take the acceleration of the universe as an example. All galaxies \emph{at all redshifts} $z$ accelerate at the same time, say $t_{\rm{ac}}\simeq7.3$ billion years after the Big Bang. With the cosmology assumed here, that corresponds to $z_{\rm{ac}}=z_2\simeq 0.67$ in the Fig. \ref{interp4}. But the only "special" thing about a galaxy at $z_2$ is that it emitted the light that we see \emph{today} just when it was on our current past light cone at the moment that the universe started to accelerate. An observer in the past, when the Hubble constant was $H=81.6\: {\rm km\:s^{-1}Mpc^{-1}}$, $\Omega_{\rm{m}}=0.49$ and $\Omega_{\rm{\Lambda}}=0.51$, and whose past light cone was the one shown with a dash-dotted line in Fig. \ref{interp4}, has already received light coming from the event "the universe started to accelerate" at $t_{\rm{ac}}\simeq7.3$ billion years. It was from objects located at  $z_{\rm{ac}}=z_1\simeq0.28$. With this example we simply want to illustrate that for a same time $t$ there exist infinite values of $z$, and that the univocal relation $t-z$ involved through Eq. \eqref{maineq} only relates the points on our past light cone.
\vspace*{2cm}

\section{Conclusions}
\label{conclusiones}

Starting with the observations made by Slipher and the pioneer works of Lema\^itre and Hubble, the cosmological redshift is one of the few observational parameters that can be measured directly. Nowadays, we think redshifts as a consequence of expansion of the universe, which is now well understood within the framework of Einstein's theory of General Relativity.

To refer to the temporal evolution of objects or cosmic structures in the universe, we often do so indistinctly through cosmic time or cosmological redshift. But we must always keep in mind that except redshift, for almost all other quantities of interest, such as recession velocities or the age of the universe, we must adopt a cosmology with certain values for the cosmological parameters. Repeatedly, the connection between cosmic time and redshift ends up generating confusion.

In this article we have shed light on some misunderstandings around the concept of the cosmological redshift, and on the $t-z$ relation in Eq. \eqref{maineq}. One should keep in mind that when we talk about different values of redshift, we are always speaking about different objects. Furthermore, at a same cosmic time there are many galaxies with different values of $z$. Snapshots labeled with different redshifts in numerical simulations are showing a temporal evolution, through the $t-z$ relation, connecting points on our past light cone, but a same comoving object in the real universe has the same cosmological redshift during its whole evolution.

\begin{acknowledgements}
G.R.B. is supported by CONICET (Argentina). G.R.B. acknowledges support from the PIP 112-2012-0100540 of CONICET (Argentina).
\end{acknowledgements}


\bibliography{bibliografia}
\bibliographystyle{apsrev}

\end{document}